\documentclass[aps,prl,twocolumn,groupedaddress,showpacs]{revtex4}
\usepackage{graphics} 
\usepackage{amsmath}
\usepackage{amsfonts}
\usepackage{amssymb}
\usepackage{bm}
\begin{document}

\title{Ultrafast control of nuclear spins using only microwave pulses: towards switchable solid-state quantum gates}

\author {George Mitrikas}
\email[Electronic address: ]{mitrikas@ims.demokritos.gr}
\author {Yiannis Sanakis}
\author {Georgios Papavassiliou}

\affiliation{Institute of Materials Science, NCSR Demokritos, 15310 Athens, Greece}

\date{\today}

\begin{abstract}

We demonstrate the control of the $\alpha$-proton nuclear spin, \textit{I}=1/2, coupled to the stable radical \textperiodcentered CH(COOH)$_{2}$, \textit{S}=1/2, in a $\gamma$-irradiated malonic acid single crystal using only microwave pulses. We show that, depending on the state of the electron spin ($m_S=\pm1/2$), the nuclear spin can be locked in a desired state or oscillate between $m_I=+1/2$ and $m_I=-1/2$ on the nanosecond time scale. This approach provides a fast and efficient way of controlling nuclear spin qubits and also enables the design of switchable spin-based quantum gates by addressing only the electron spin.

\end{abstract}

\pacs{03.67.Lx, 76.30.-v, 76.60.-k}
\maketitle

Since the idea of quantum information processing (QIP) fascinated the scientific community \cite{ref1,ref2,ref3}, electron and nuclear spins have been regarded as promising candidates for quantum bits (qubits) \cite{ref4}. Nuclear magnetic resonance (NMR) in liquid state was the first spectroscopic technique used to demonstrate several quantum computation algorithms \cite{ref13,ref14,ref15,ref16}, while the emerged scalability limitations can be overcome by using electron paramagnetic resonance (EPR) spectroscopy \cite{ref17}. The construction of quantum gates based exclusively on electron spins with controllable exchange interactions is however a challenging task, especially when crystalline materials of this kind are on demand. This difficulty has motivated an on-going effort to find appropriate electron spin qubits, including for instance quantum dots \cite{ref18}, single-molecule magnets \cite{ref19}, and antiferromagnetic heterometallic rings \cite{ref20} or metal clusters \cite{ref21}. On the other hand, hybrid electron-nuclear spin systems with long decoherence times can be found in a variety of materials like organic single crystals \cite{ref5}, endohedral fullerenes \cite{ref6,ref7,ref8}, phosphorous donors in silicon crystals \cite{ref9}, and nitrogen-vacancy centres in diamond \cite{ref10,ref11}. They have been used to perform two-qubit quantum operations, demonstrate entangled states, or build solid-state quantum memories. These systems benefit from well-defined and separated EPR and NMR transitions that can be selectively manipulated by resonant mw and radio frequency (rf) radiation, respectively. Therefore, advanced EPR methods \cite{ref22} employing both selective mw and rf pulses play a key role in QIP based on electron-nuclear spin systems.

The significant difference between decoherence times of electron (\textit{T}$_{2e}$) and nuclear (\textit{T}$_{2n}$) spins, which is due to the fact that their gyromagnetic ratios $\gamma_i$ differ by two to three orders of magnitude, has been utilized for the realization of a quantum memory that uses slow relaxing nuclear spins to store the information and fast relaxing electron spins for processing and readout \cite{ref9}. On the other hand, this difference also restricts the clock rates of gate operation to the corresponding ones of the slow relaxing qubit, namely to the MHz frequency scale. More seriously though, the combination of the small gyromagnetic ratios $\gamma_n$ of nuclei with the currently available rf fields $B_1$, results in slow rotations of nuclear spins. For instance, the nutation (Rabi) frequency of a proton nuclear spin for a typical value of $B_1$=1 mT is $\omega_1/2\pi=\gamma_n B_1/h=$42.6 kHz which implies a length of $\Delta t_{\pi}=11.7 \mu$s for an rf $\pi$-pulse. This time interval is of the same order of magnitude with typical electron decoherence times, \textit{T}$_{2e}$, and thus the efficiency of the two-qubit quantum gate may become questionable due to relaxation losses.

One way to overcome this difficulty is to utilize the hyperfine interaction which is typically stronger than the Rabi nuclear frequency, i.e. of the order of some MHz in organic radicals.
\begin{figure}[b]
\includegraphics{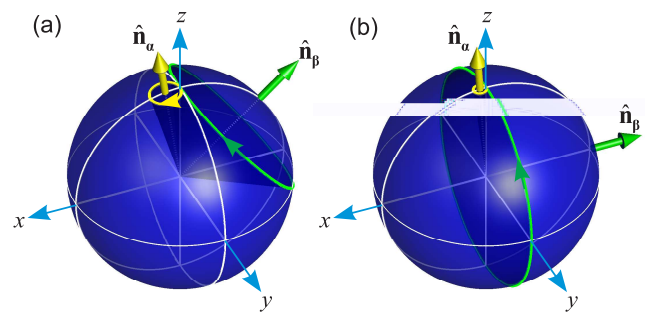}
\caption{\label{fig1}(Color online) (a) Bloch sphere showing the trajectories of nuclear spin magnetization under free evolution assuming that the nuclear spin is polarized in the beginning ($m_I$=+1/2). The quantization axis of the effective magnetic field experienced by the nucleus depends on the state of the electron spin: for $m_S$=+1/2 the nuclear spin precesses about $\bm{\hat{n}_{\alpha}}$ (yellow trace) with angular frequency $\omega_{\alpha}=|\omega_{12}|$; for $m_S$=-1/2 the nuclear spin precesses about $\bm{\hat{n}_{\beta}} $(green trace) with angular frequency $\omega_{\beta}=|\omega_{34}|$. (b) Exact cancellation case, $\omega_I=A/2$.}
\end{figure}
When this interaction is anisotropic the quantization axes of the hyperfine fields deviate from the \textit{z} axis (see FIG. \ref{fig1}a) and thus the nutating nuclear spin can be inverted by implementing a proper set of mw $\pi$-pulses separated by suitable time delays \cite{ref23}. This concept has been recently used in a similar fashion \cite{ref24} to demonstrate the creation of nuclear spin coherence via amplitude-modulated mw pulses \cite{ref30}.

Herein we focus on a special case of electron spin echo envelope modulation (ESEEM) spectroscopy, the so-called exact cancellation \cite{ref22} where the hyperfine and electron Zeeman interaction cancel each other in one $m_S$ manifold. FIG. \ref{fig1}b shows that in this case the effective magnetic field at the nucleus is either almost parallel or exactly perpendicular to the \textit{z} axis, depending on the state of the electron spin. Consequently, during free evolution of the system with $m_S=-1/2$, the nuclear spin naturally oscillates between the $m_I=+1/2$ and $m_I=-1/2$ states without the need of rf field. In addition, it can be locked in a desired state by inverting the electron spin state to $m_S$=+1/2 with a mw $\pi$-pulse applied after a proper time delay.

\begin{figure}
\includegraphics{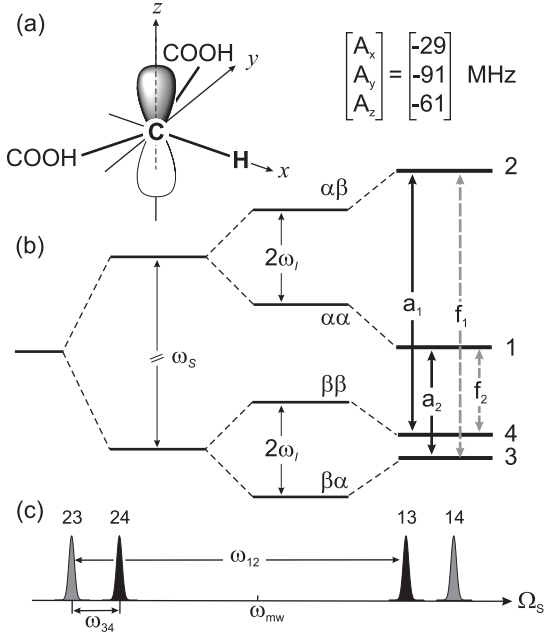}
\caption{\label{fig2} (a) Schematic representation of the stable radical \textperiodcentered CH(COOH)$_{2}$ and the principal axes system of the hyperfine tensor. (b) Energy level diagram depicting the allowed (a1, a2) and forbidden (f1, f2) electron spin transitions. (c) EPR stick spectrum.}
\end{figure}

To demonstrate this control, a single crystal of  $\gamma$-irradiated malonic acid is used. Here the paramagnetic species is the stable radical \textperiodcentered CH(COOH)$_2$ shown in  FIG. \ref{fig2}a, where the unpaired electron (\textit{S}=1/2) resides on the carbon 2\textit{p}$_z$ orbital and is hyperfine-coupled to the $\alpha$-proton nuclear spin (\textit{I}=1/2)\cite{ref12}. The rotating frame spin Hamiltonian is given by \cite{ref22}
\begin{equation}
{\cal H}_0={\Omega}_SS_z+{\omega}_II_z+AS_zI_z+BS_zI_x
\label{eq1}
\end{equation}
where $\Omega_S=\omega_S-\omega_{mw}$ is the offset of the electron Zeeman frequency $\omega_S=g\beta_e\textit{B}_0/\hbar$ from the mw frequency $\omega_{mw}$, $\omega_I=-g_n\beta_n\textit{B}_0/\hbar$ is the nuclear Zeeman frequency, $g$ and $g_n$ are the electron and nuclear g-factors, $\beta_e$ and $\beta_n$ are the Bohr and nuclear magnetons, $B_0$ is the static magnetic field along \textit{z}-axis, and \textit{A}, \textit{B} describe the secular and pseudo-secular part of the hyperfine coupling. In this four-level electron-nuclear spin system (FIG. \ref{fig2}b) there are six possible transitions: four EPR with $\Delta m_S=\pm 1$ as illustrated in the stick spectrum of FIG. \ref{fig2}c, and two NMR
\begin{figure*}
\includegraphics{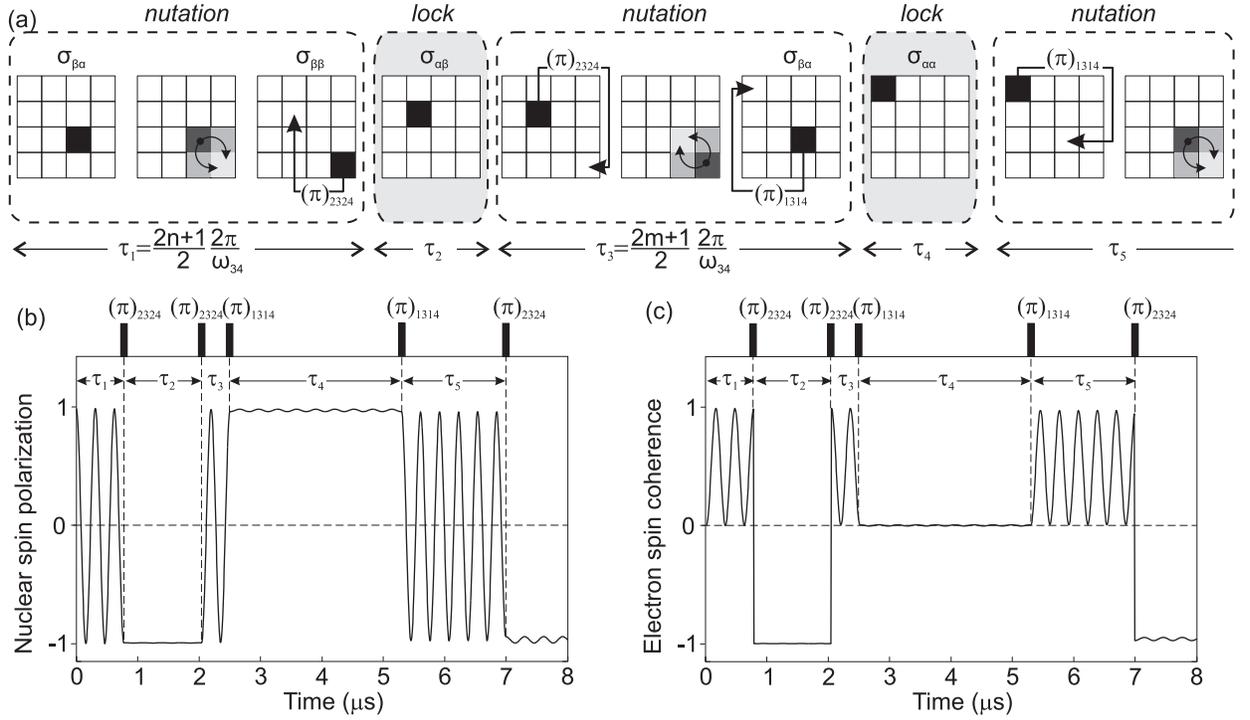}
\caption{\label{fig3} (a) Representation of the density matrix in the product basis $|\alpha\alpha\rangle$, $|\alpha\beta\rangle$, $|\beta\alpha\rangle$, $|\beta\beta\rangle$ during nutation or locking time periods. (b) Numerical simulation of the nuclear spin polarization $\langle I_z\rangle$ during the applied pulse sequence that uses semi-selective mw $\pi$-pulses (shown at the top). Spin Hamiltonian parameters: nuclear Zeeman frequency, $\omega_I/2\pi=-14.58$ MHz; hyperfine coupling constants, $A/2\pi=-29.06$ MHz, $B/2\pi=6.45$ MHz. (c) Corresponding numerical simulation of the electron spin coherence $\langle S_x\rangle$ after the detection pulse sequence $(\pi/2)_{2324}-\tau-(\pi)_{2324}-\tau-$ echo with $\tau=2\pi/\omega_{34}=$310 ns.}
\end{figure*}
with frequencies $\omega_{12}$ and $\omega_{34}$. For a proton nuclear spin ($\omega_I<0$) and a negative hyperfine coupling ($A<0$) in the weak-coupling regime ($|A|<2|\omega_I|$), the singed nuclear transition frequencies can be expressed as $\omega_{12}=-\sqrt{(\omega_I+A/2)^2+(B/2)^2}$ and $\omega_{34}=-\sqrt{(\omega_I-A/2)^2+(B/2)^2}$. In addition, the quantization axes of the effective frequency vectors are determined by the angles $\eta_{\alpha}=\arctan[-B/(A+2\omega_I)]$ and $\eta_{\beta}=\arctan[-B/(A-2\omega_I)]$ which are defined relative to \textit{z} direction. For the case of interest of exact cancellation, $A=2\omega_I$, the nuclear transition frequencies become $\omega_{12}=-\sqrt{(2\omega_I)^2+(B/2)^2}$ and $\omega_{34}=-B/2$   (with $B>0$) with the corresponding angles being $\eta_{\alpha}=\arctan[-B/4\omega_I]$ and $\eta_{\beta}=-\pi/2$.

The concept of manipulating the proton nuclear spin solely by using mw pulses can be best described in terms of the density matrix formalism. Figure \ref{fig3}a shows an example of this control assuming that the system is initially prepared in the pseudopure state $\sigma_{\beta\alpha}\equiv\ |\beta\alpha\rangle\langle\beta\alpha |$ which is described by the density matrix in the product basis
\[ \sigma_{\beta\alpha}=\left( \begin{array}{cccc}
0 & 0 & 0 & 0 \\
0 & 0 & 0 & 0 \\
0 & 0 & 1 & 0 \\
0 & 0 & 0 & 0\end{array} \right)\]
\\During free evolution of the system under the spin Hamiltonian (Eq. \ref{eq1}) the hyperfine field drives the nuclear spin transition and the population is periodically transferred from $\sigma_{\beta\alpha}$ to $\sigma_{\beta\beta}$ and vice-versa with frequency $\omega_{34}$. This transfer is complete (i.e. $\sigma_{\beta\beta}\equiv\ |\beta\beta\rangle\langle\beta\beta |$) only if the exact cancellation condition $\sin\eta_{\beta}=-1$ is fulfilled, and occurs for free evolution times $\tau_1=(2m+1)\pi/\omega_{34},(m=0,1,2,...$) (see supplementary information). A subsequent semi-selective mw  $\pi$-pulse, that simultaneously excites the forbidden 23 and allowed 24 EPR transitions \cite{ref25,ref26} (represented as $(\pi)_{2324}$), transfers the population to the $|\alpha\beta\rangle$ state. Provided that the angle $\eta_{\alpha}$ is very small, the populations and nuclear coherences in the $m_S=+1/2$ subspace are virtually trapped because of the negligible branching between $|\alpha\alpha\rangle$ and $|\alpha\beta\rangle$ states. Under that treatment the nuclear spin can be locked in the $m_I=-1/2$ state for arbitrary time, $\tau_2$.
An additional $(\pi)_{2324}$-pulse will transfer back the population to the $|\beta\beta\rangle$ state and the nuclear spin will start oscillating again between $m_I=+1/2$ and $m_I=-1/2$ with frequency $\omega_{34}$. Quantitatively, this is demonstrated in FIG. \ref{fig3}b that shows the simulation of the nuclear spin polarization, $\langle I_z\rangle$.
\begin{figure}
\includegraphics{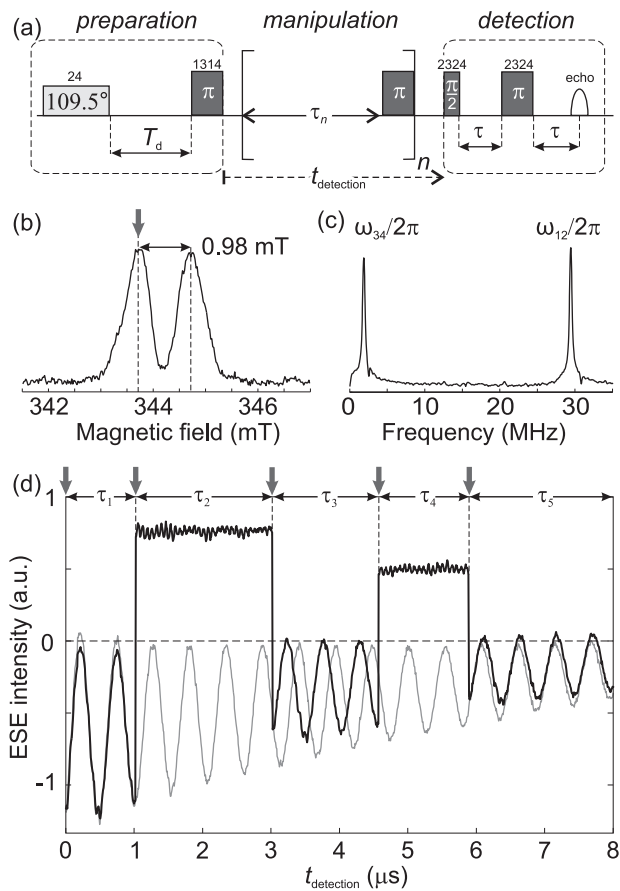}
\caption{\label{fig4} (a) The proposed pulse sequence consists of three parts: preparation, for creating the pseudopure state $\sigma_{\beta\alpha}$; manipulation, consisting of a time delay $\tau_n$ and a semi-selective $(\pi)_{2324}$ or $(\pi)_{1314}$ pulse; detection of electron spin coherence through a two-pulse echo with $\tau=2\pi/\omega_{34}$. (b) Free induction decay (FID)-detected EPR spectrum. The arrow indicates the observer position $B_0=343.7$ mT of the semi-selective $(\pi)_{2324}$ pulses. (c) Three-pulse ESEEM spectrum. (d) Electron spin echo (ESE) intensity as a function of time $t_{\text{detection}}$ between preparation and detection.}
\end{figure}
Therefore, by choosing appropriate time delays and semi-selective mw pulses, the full control of the nuclear spin can be achieved.

Although direct measurement of nuclear spin polarization is not possible in EPR spectroscopy, FIG. \ref{fig3}c shows that the electron spin coherence $\langle S_x\rangle$ after the sequence $(\pi/2)_{2324}-\tau-(\pi)_{2324}-\tau-$echo with $\tau=2\pi m/\omega_{34}, (m=1,2,3...)$ represents the population difference between states 2 and 4. Consequently, such a sub-sequence can be used to monitor the state of our system.

For the experimental demonstration of this control the system can start from a pseudopure state, e.g. $\sigma_{\beta\alpha}$. For the case of isotropic hyperfine coupling ($B=0$), such a state can be prepared with a selective mw pulse $P_{24}(\beta_0)$, with $\beta_0=\arctan(-1/3)=109.5^{\circ}$, followed by a $\pi/2$-rf pulse, $P_{12}(\pi/2)$\cite{ref4}. For anisotropic hyperfine coupling and a strong mixing of states the above pulse sequence does not result in the desired pseudopure state. However, for the special case of exact cancellation, the operation of a selective mw pulse $P_{24}(\beta)$ (with effective rotation angle $\beta_{eff}=\arctan(-1/3)=109.5^{\circ}$) at the thermal equilibrium density matrix, $\sigma_0=-S_z$, gives the pseudopure state $\sigma_{\alpha\alpha}$ in good approximation (see supplementary information). An additional semi-selective $(\pi)_{1314}$ pulse applied after time $T_d>5T_{2e}$ complements the preparation part of the pulse scheme shown in FIG. \ref{fig4}a and leads the system to the pseudopure state $\sigma_{\beta\alpha}$.

The experiments were performed at a crystal orientation for which the hyperfine coupling is very close to exact cancellation: the three-pulse ESEEM spectrum (FIG. \ref{fig4}c) measured at the observer position $B_0=343.7$ mT $(\omega_I/2\pi=-14.6$ MHz) shows two peaks at frequencies $\omega_{12}/2\pi=-29.42$ MHz and $\omega_{34}/2\pi=-1.83$ MHz from which the angles $\eta_{\alpha}=-3.6^{\circ}$ and $\eta_{\beta}=-87.1^{\circ}$ can be inferred. Figure \ref{fig4}b shows that the two EPR transitions 23 and 24 are not resolved due to the small nuclear frequency $\omega_{34}$ and inhomogeneous broadening. Therefore, an accurate excitation of transition 24 with a selective $P_{24}$ pulse is difficult to be accomplished. Moreover, the necessity of the semi-selective $(\pi)_{1314}$ pulse requires an additional mw frequency which was not available with the current experimental setup. However, both technical issues were circumvented by replacing the preparation part of FIG. \ref{fig4}a with a semi-selective $(\pi)_{2324}$ pulse that interchanges populations between levels 2 and 4. Their population difference (which is our observable here) is expected to oscillate between zero and a maximum value because after the semi-selective pulse levels 3 and 4 will start exchanging populations, in analogy to the situation of population transfer during free evolution of $\sigma_{\beta\alpha}$ under the spin Hamiltonian (Eq. \ref{eq1}). Figure \ref{fig4}d (gray trace) shows that the signal virtually vanishes for $t=(m+1/2)T$ $(m=0,1,2,...)$, where $T=504$ ns ($\approx 2\pi/\omega_{34}$), implying that our system is indeed very close to exact cancellation. By applying semi-selective mw pulses, the nuclear spin evolution can be locked in the $m_I=-1/2$ state or released at will (FIG. \ref{fig4}d, black trace).

In conclusion, we have demonstrated that our approach provides a fast way of controlling nuclear spins in hybrid spin systems and thus overcomes the problem of relaxation incompatibility between electron and nuclear spins. Furthermore, the method offers a means for controlling the interaction between the two qubits in a fast and efficient way. Although the hyperfine interaction is always present, the special case of exact cancellation considered here induces two different effective states of interaction regarding the behaviour of the nuclear spin: an active (nutation) and a passive state (lock) that can be switched with a single mw pulse. While the control of dipolar interactions between neighbouring electron spin qubits has been proposed using SWAP operations \cite{ref27,ref28}, to the best of our knowledge to date this kind of control in similar spin systems has been experimentally demonstrated only once using decoupling methods \cite{ref29}.

Our work shows that advanced pulsed EPR methods like ESEEM spectroscopy can play a key role in QIP based on hybrid electron-nuclear spin systems. Furthermore, it gives new perspective to such systems by considering them not only for performing quantum memories but also for building solid-state quantum gates. Given that the development of new materials as well as new methods is equally important for the realization of a quantum computer, it is likely that the scheme presented here will become a standard method for manipulating electron and nuclear spin qubits and controlling their hyperfine interaction in the future real quantum computers.

\end{document}


\begin{center}
\textbf{SUPPLEMENTARY INFORMATION}

for the manuscript

\textbf{Ultrafast control of nuclear spins using only microwave pulses: towards switchable solid-state quantum gates}

by

George Mitrikas, Yiannis Sanakis, and Georgios Papavassiliou

\emph{Institute of Materials Science, NCSR Demokritos, 15310 Athens, Greece}

email: mitrikas@ims.demokritos.gr
\end{center}

\section{\label{sec:level1} Theory}
\subsection{\label{sec:level2} Population transfer between $|\beta\alpha\rangle$ and $|\beta\beta\rangle$ as driven by the hyperfine field.}

The pseudopure state $\sigma_{\beta\alpha}$ is described by the density matrix in the Cartesian product basis (CPB)
\begin{equation}
\sigma_{\beta\alpha}=\left( \begin{array}{cccc}
0 & 0 & 0 & 0 \\
0 & 0 & 0 & 0 \\
0 & 0 & 1 & 0 \\
0 & 0 & 0 & 0\end{array} \right)
\label{eq1}
\end{equation}
The spin Hamiltonian
${\cal H}_0={\Omega}_SS_z+{\omega}_II_z+AS_zI_z+BS_zI_x$
is diagonalized (${\cal H}_0^d=U{\cal H}_0U^{\dagger}$)
by the unitary transformation
\begin{equation}
U=\left( \begin{array}{cccc}
\cos(\frac{\eta_{\alpha}}{2}) & -\sin(\frac{\eta_{\alpha}}{2}) & 0 & 0 \\
\sin(\frac{\eta_{\alpha}}{2}) & \cos(\frac{\eta_{\alpha}}{2}) & 0 & 0 \\
0 & 0 & \cos(\frac{\eta_{\beta}}{2}) & -\sin(\frac{\eta_{\beta}}{2}) \\
0 & 0 & \sin(\frac{\eta_{\beta}}{2}) & \cos(\frac{\eta_{\beta}}{2})\end{array} \right)
\label{eq2}
\end{equation}
where $\eta_{\alpha}=\arctan[-B/(A+2\omega_I)]$, $\eta_{\beta}=\arctan[-B/(A-2\omega_I)]$,
and $\eta=(\eta_{\alpha}-\eta_{\beta})/2.$
Therefore, the initial density matrix $\sigma_1=\sigma_{\beta\alpha}$ is expressed in the \textit{eigenbasis} as
\begin{equation}
\sigma_1'=U\sigma_1U^{\dagger}=\left( \begin{array}{cccc}
0 & 0 & 0 & 0 \\
0 & 0 & 0 & 0 \\
0 & 0 & \cos^2(\frac{\eta_{\beta}}{2}) & \sin(\frac{\eta_{\beta}}{2})\cos(\frac{\eta_{\beta}}{2}) \\
0 & 0 & \sin(\frac{\eta_{\beta}}{2})\cos(\frac{\eta_{\beta}}{2}) & \sin^2(\frac{\eta_{\beta}}{2})\end{array} \right)
\label{eq3}
\end{equation}
which, after free evolution under the spin Hamiltonian ${\cal H}_0^d$ for time $t$, becomes
\begin{equation}
\sigma_t'=\exp(-i{\cal H}_0^dt)\sigma_1'\exp(i{\cal H}_0^dt)=\left( \begin{array}{cccc}
0 & 0 & 0 & 0 \\
0 & 0 & 0 & 0 \\
0 & 0 & \cos^2(\frac{\eta_{\beta}}{2}) & \sin(\frac{\eta_{\beta}}{2})\cos(\frac{\eta_{\beta}}{2})e^{-i\omega_{34}t} \\
0 & 0 & \sin(\frac{\eta_{\beta}}{2})\cos(\frac{\eta_{\beta}}{2})e^{i\omega_{34}t} & \sin^2(\frac{\eta_{\beta}}{2})\end{array} \right)
\label{eq4}
\end{equation}
where $\omega_{34}=-\sqrt{(\omega_I-A/2)^2+(B/2)^2}$ is the signed nuclear transition frequency in the  $\beta$-electron spin manifold. This density matrix in \textit{CPB} reads
\begin{equation}
\sigma_t=U^{\dagger}\sigma_t'U=\left( \begin{array}{cccc}
0 & 0 & 0 & 0 \\
0 & 0 & 0 & 0 \\
0 & 0 & 1-\frac{1}{2}\sin^2(\eta_{\beta})[1-\cos(\omega_{34}t)] & \begin{split}-\frac{1}{2}\sin(2\eta_{\beta})\sin^2(\frac{\omega_{34}t}{2})-\\ \frac{i}{2}\sin(\eta_{\beta})\sin(\omega_{34}t) \end{split} \\
0 & 0 & \begin{split}-\frac{1}{2}\sin(2\eta_{\beta})\sin^2(\frac{\omega_{34}t}{2})+ \\ \frac{i}{2}\sin(\eta_{\beta})\sin(\omega_{34}t) \end{split} & \frac{1}{2}\sin^2(\eta_{\beta})[1-\cos(\omega_{34}t)]\end{array} \right)
\label{eq5}
\end{equation}
From Eq. \ref{eq5} we see that populations and nuclear coherences are periodic functions of time with angular frequency $\omega_{34}$. The initial state $\sigma_1=\sigma_{\beta\alpha}$ recurs for times $T=2m\pi/\omega_{34}$ $(m=1,2,3,...)$. At exact cancellation ($\sin(\eta_{\beta})=-1$ ) and times $\tau=(2m+1)\pi/\omega_{34}$ $(m=1,2,3,...)$ complete population transfer occurs, i.e. $\sigma_{\tau}=\sigma_{\beta\beta}$.

\subsection{\label{sec:level3} Preparation of the pseudopure state $\sigma_{\beta\alpha}$.}

Starting from the thermal equilibrium density matrix $\sigma_0=-S_z$, the standard method for preparing the pseudopure state $\sigma_{\beta\alpha}$ is the application of a selective mw pulse $P_{24}(\beta_0)$, with nominal rotation angle $\beta_0=\arctan(-1/3)=109.5^{\circ}$, followed by a selective rf pulse, $P_{12}(\pi/2)$.  This is true only for isotropic hyperfine coupling, where $\eta=0$.  Here we examine the case of a non-zero anisotropic hyperfine coupling and particularly the case of exact cancellation. A selective mw pulse along the rotating frame y-axis, which is on-resonance with the allowed EPR transition 24 and has a nominal rotation angle $\beta$, is given by
\begin{equation}
P_{24}(\beta)=\exp[-i\beta\cos{\eta}S_y^{24}], \text{ with } S_y^{24}=\frac{1}{2}S_y-S_yI_z
\label{eq6}
\end{equation}
Therefore, after this pulse the density matrix is expressed in the \textit{eigenbasis} as
\begin{equation}
\sigma'=P_{24}(\beta)\sigma_0P_{24}^{-1}(\beta)=\left( \begin{array}{cccc}
-\frac{1}{2} & 0 & 0 & 0 \\
0 & -\frac{1}{2}\cos2\phi & 0 & -\frac{1}{2}\sin2\phi \\
0 & 0 & \frac{1}{2} & 0 \\
0 & -\frac{1}{2}\sin2\phi & 0 & \frac{1}{2}\cos2\phi\end{array} \right)
\label{eq7}
\end{equation}
with $2\phi=\beta\cos\eta$. When the condition of exact cancellation is fulfilled ($\sin(\eta_{\beta})=-1$ ), the transformation back to \textit{CPB} gives
\begin{equation}
\sigma=U^{\dagger}\sigma'U=\left( \begin{array}{cccc}
\begin{split}\text{A}=-\frac{1}{2}\cos^2(\frac{\eta_{\alpha}}{2})-\\ \frac{1}{2}\cos2\phi\sin^2(\frac{\eta_{\alpha}}{2}) \end{split} &  &  &  \\
 & \begin{split} \text{B}=\frac{1}{2}\sin^2(\frac{\eta_{\alpha}}{2})-\\ \frac{1}{2}\cos2\phi\cos^2(\frac{\eta_{\alpha}}{2}) \end{split}&  &  \\
 &  & \text{C}=\frac{1}{4}+\frac{1}{4}\cos2\phi &  \\
 &  &  & \text{D}=\frac{1}{4}+\frac{1}{4}\cos2\phi \end{array} \right)
\label{eq8}
\end{equation}
For clarity, the off-diagonal elements of the density matrix in Eq. \ref{eq8} have been omitted. They represent electron and nuclear spin coherences that have completely decayed after time $T_d>5T_{2e}$. We can now determine the angle $\phi=\frac{\beta}{2}\cos\eta$ in order to fulfil the condition B=C=D, i.e. to get the pseudopure state $\sigma_{\alpha\alpha}$. From this requirement we find
\[ \frac{1}{2}\sin^2(\frac{\eta_{\alpha}}{2})-\frac{1}{2}\cos2\phi\cos^2(\frac{\eta_{\alpha}}{2})=\frac{1}{4}+\frac{1}{4}\cos2\phi \]
which gives
\begin{equation}
\cos2\phi=\frac{-1+2\sin^2(\frac{\eta_{\alpha}}{2})}{2\cos^2(\frac{\eta_{\alpha}}{2})+1}\approx-\frac{1}{3}+\frac{\eta^2_{\alpha}}{6}
\label{eq9}
\end{equation}
The approximation in Eq. \ref{eq9} holds for small angles $\eta_{\alpha}=\arctan(-B/4\omega_I)$, i.e. when $B\ll4\omega_I$. Therefore, application of a selective mw pulse $P_{24}(\beta)$ with effective rotation angle $\beta_{eff}=\beta\cos\eta=\arctan(-1/3)=109.5^{\circ}$ at the thermal equilibrium density matrix, gives in good approximation the pseudopure state $\sigma_{\alpha\alpha}$ without the need of rf pulse. An additional semi-selective $(\pi)_{1314}$ pulse applied after time $T_d$ drives the system to the pseudopure state $\sigma_{\beta\alpha}$.

\section{\label{sec:level4} Materials and methods}

Single crystals of malonic acid were grown from saturated aqueous solution by slow evaporation and then irradiated at room temperature with a $^{60}$Co $\gamma$-ray source. The occurring paramagnetic species is the stable radical \textperiodcentered CH(COOH)$_{2}$ shown in  Fig. 2a, where the unpaired electron ($S=1/2$) resides on the carbon 2$p_z$ orbital and is hyperfine-coupled to the $\alpha$-proton nuclear spin ($I=1/2$).  The hyperfine interaction is highly anisotropic and exhibits rhombic symmetry with principal values $(A_x,A_y,A_z)=(-29,-91,-61)$ MHz.

Pulsed EPR measurements were performed at room temperature using a Bruker ESP380E spectrometer operating at X-band frequency ($\omega_{mw}/2\pi=9.646$ GHz). The field-swept EPR spectrum of Fig. 4b was recorded by integrating the FID after a selective mw $\pi/2$ pulse of 500 ns. The three-pulse ESEEM spectrum was obtained with the pulse sequence $\pi/2-\tau-\pi/2-T-\pi/2-$echo, with a $\pi/2$ pulse of length 16 ns and an interpulse delay $\tau=200$ ns. Time $T$ was increased in steps of d$t=8$ ns starting from $T=56$ ns (512 points). A four-step phase cycling was used to remove unwanted echoes. The time trace was baseline corrected with a two-order exponential, apodized with a Gaussian window, and zero-filled. After Fourier transform the absolute-value spectrum was calculated (Fig. 4c).

Our experiments for the demonstration of the control of the proton nuclear spin were performed on the low field line of the EPR doublet as indicated by the arrow of Fig. 4b. $\pi$-pulses of length 32 ns were used corresponding to mw field strength of $\omega_1/2\pi=15.6$ MHz. This is smaller than the hyperfine splitting at this crystal orientation ($\approx$29 MHz) and therefore the pulse excites only transitions 23 and 24 (which are separated by $\omega_{34}/2\pi=-1.83$ MHz). Such a pulse is called semi-selective and is represented as $(\pi)_{2324}$. The state of the system was monitored by measuring population differences with the pulse sequence $(\pi/2)_{2324}-\tau-(\pi)_{2324}-\tau-$echo that uses $\tau=552$ ns. A four-step phase cycling was applied to remove unwanted echoes created by the semi-selective $(\pi)_{2324}$-pulses of the manipulation part of the control scheme. In Fig. 4d the preparation part of the pulse sequence is replaced by a semi-selective $(\pi)_{2324}$ pulse. After this pulse the signal oscillates with frequency $\omega_{\beta}/2\pi=1.83$ MHz under free evolution (gray trace). By applying semi-selective $(\pi)_{2324}$ pulses at positions indicated by arrows the nuclear spin can be locked in the $m_I=-1/2$ state (black trace, locking time intervals $\tau_2=2016$ ns and $\tau_4=1288$ ns) and then released at any point (black trace, nutation time intervals $\tau_3=1504$ ns and $\tau_5=2096$ ns).